\documentclass[aps,prl,twocolumn,preprintnumbers,10pt,showpacs]{revtex4-1}
\usepackage{amsmath,bm}

\usepackage{graphicx}

\allowdisplaybreaks

\begin{document}

\preprint{MPP-2013-236, ZU-TH 16/13}

\title{Precise QCD predictions for the production\\ of a photon pair in 
association with two jets}

\author{T.\ Gehrmann$^a$, N.\ Greiner$^b$, G.\ Heinrich$^b$}

\affiliation{$^a$ Institute for Theoretical Physics, University of
  Z\"urich, CH-8057 Z\"urich, Switzerland\\
$^b$ Max Planck Institut f\"ur Physik, F\"ohringer Ring 6, D-80805 M\"unchen, Germany}


\begin{abstract}
  We compute the cross section for the production of a high-mass photon pair 
  in association with two hadronic jets to next-to-leading order (NLO) in 
  quantum chromodynamics (QCD). Our results allow for the first time to reliably predict the 
  absolute normalisation of this process, and demonstrate that the shape of important kinematical 
  distributions is modified by higher-order effects.    
  The perturbative corrections will be an important ingredient to precision studies of 
  Higgs boson properties from its production in association with two jets.  
  
\end{abstract}

\maketitle

Following the discovery of the Higgs boson at the CERN LHC~\cite{higgsLHC}, the study of 
Higgs boson properties has now become a major research objective of particle physics. 
By measuring a variety of production and decay modes of the Higgs boson, the determination of 
the Higgs boson quantum numbers and 
couplings to Standard Model particles will become increasingly precise, thereby 
allowing to uncover possible deviations from the Standard Model realization of the Higgs 
mechanism of electroweak symmetry breaking. These studies rely on a close interplay between 
experimental data and theoretical predictions for Higgs boson signal and background processes. 

Reliable theoretical predictions for 
hadron collider processes require the inclusion of higher order QCD corrections.  
Impressive progress has been made in recent years in the derivation of QCD corrections to 
the most important Higgs boson production processes, with gluon 
fusion~\cite{ggHnnlo} and associated production~\cite{vhnnlo} known 
fully exclusively to next-to-next-to-leading 
order (NNLO) in QCD, and vector-boson-fusion~\cite{vbfnlo} and associated production 
with top quarks~\cite{ttHnlo} to next-to-leading order (NLO).  
For the gluon fusion process, which is the largest  contribution to Higgs production 
at the LHC, NLO corrections 
have also been derived for Higgs boson production in association with up to 
three jets~\cite{ggH1jnlo,ggH2jnlo,ggH3jnlo}. 

To disentangle different Higgs boson production processes, and to optimize signal-to-background 
ratios, one often distinguishes samples according to the number of jets observed 
together with the Higgs boson candidate. In particular, Higgs production in association with 
two jets allows to probe the vector boson fusion process, which is of crucial importance 
for the study of the Higgs mechanism and for the determination of couplings. 
Accordingly, precise predictions for the corresponding 
background processes are required to optimize detection strategies and to allow for a meaningful
interpretation of experimental observations. 

Among the most prominent Higgs boson discovery 
modes is the decay to two photons $H\to \gamma\gamma$~\cite{haa}, 
which, despite its low branching fraction, has a 
favourable signal-to-background ratio. In the experimental studies of 
Higgs boson properties in this channel~\cite{atlashaa,cmshaa}, the 
background contributions are estimated from fits to side-band data with 
diphoton invariant mass away from the Higgs boson mass. This 
pragmatic approach to quantify the total background allowed the Higgs boson 
discovery in the diphoton channel; it may however face its limitations once it comes to 
precision studies of the Higgs boson properties, in particular its production mechanism. In this 
context, it is highly desirable to have precise predictions for the production of photon pairs 
in association with a definite number of hadronic jets. Photon pair production without extra 
jets is known to NNLO~\cite{ggaannlo}, and photon-pair-plus-one-jet production to 
NLO~\cite{nagy,aa1jnlo}. In both cases, the inclusion of higher order corrections revealed new 
kinematical features that may turn out to be crucial in precision studies. For photon-pair-plus-two-jet 
production, only leading order predictions were available up to now, 
which are insufficient for precise phenomenological studies. 
In this letter, we present the first calculation of 
NLO QCD corrections to this process. 

Photons at hadron colliders can originate either from the hard interaction process itself or from 
hadron decays. 
To single out the photons originating from the hard production process, 
photon isolation criteria are applied, 
which are typically formulated in the form of a maximum amount of hadronic energy that is 
admitted in the vicinity of the photon. 
By admitting some hadronic activity around a photon, one includes final state configurations 
with a final state quark radiating a highly energetic collinear photon. These configurations contain 
a collinear singularity, related to small invariant masses of the quark-photon system. Mass 
factorization in QCD relates this singularity to a redefinition of the quark-to-photon 
fragmentation function~\cite{walsh,morgan}, 
which describes the production of a photon inside a hadronic jet. 
Like parton distributions in the proton, these fragmentation functions are non-perturbative 
objects that have to be determined from experimental data~\cite{aleph}. 

To suppress the dependence of isolated photon cross sections on these a priori unknown 
fragmentation functions, a smooth
cone isolation criterion has been proposed~\cite{frixione}, which 
varies the 
 threshold on the hadronic energy inside the isolation cone with the 
 radial distance from the photon. It is described by the cone size $R$, 
 a weight factor $n$ and 
 an isolation parameter $\epsilon$. With this criterion, one considers smaller 
 cones of radius $r_\gamma$ inside the $R$-cone and calls the photon isolated 
 if the energy in any sub-cone does not exceed
 \begin{displaymath}
 E_{{\rm had, max}} (r_{\gamma}) = \epsilon \,p_{T}^{\gamma} \left( \frac{1-\cos r_\gamma}
 {1-\cos R}\right)^{n}\;.
 \label{eq:frix}
 \end{displaymath}
By construction, the smooth cone isolation 
 does not admit any hard collinear quark-photon configurations, thereby allowing a full 
 separation of 
 direct and secondary photon production, and consequently eliminating the need for 
 a photon fragmentation contribution in the theoretical description. 
 We will employ the smooth isolation criterion throughout our calculation.

For the production of two photons and two jets the calculation of 
three types of subprocesses  is needed, namely
\begin{displaymath}
 q \bar{q} \to \gamma \gamma q \bar{q} , \quad
 q \bar{q} \to \gamma \gamma q' \bar{q}',\quad
 g g \to \gamma \gamma q \bar{q} \;.
\end{displaymath}
All other subprocesses can be obtained by crossing and/or changing of overall prefactors. 
We neglect the contributions from the loop-suppressed process $gg \to \gamma\gamma gg$,
which is formally of higher order in the perturbative expansion. To obtain the NLO 
predictions for each subprocess, virtual one-loop corrections and single real radiation 
corrections have to be computed. 

The matrix elements for tree level and real emission contributions have been generated with \texttt{MadGraph}~\cite{mg4},
the subtraction terms to cancel the QCD singularities are provided by 
\texttt{MadDipole}~\cite{maddipole},
which uses the dipole formalism as described in \cite{Catani:1996vz}.
The tree level and IR subtracted NLO real radiation cross sections for diphoton plus two jet 
final states have been checked 
against \texttt{SHERPA}~\cite{sherpa}, finding good agreement.
For the generation of the virtual one-loop amplitudes the 
package \texttt{GoSam}~\cite{gosam} has been used.
Based on a Feynman diagrammatic approach, it uses 
\texttt{QGRAF}~\cite{qgraf} and \texttt{FORM}~\cite{form}
to generate the diagrams. Furthermore it uses the library 
\texttt{Spinney}~\cite{spinney} to deal with
the spinor-helicity formalism and \texttt{Haggies}~\cite{haggies} 
and \texttt{FORM} to optimise the output.
For the reduction we use a $d$-dimensional integrand level 
decomposition as implemented in \texttt{Samurai}~\cite{samurai},
applying unitarity based methods ~\cite{unitarity}. 
For unstable points we used a tensorial
decomposition as contained in  \texttt{Golem95}~\cite{golem95}. 
The remaining master integrals are computed with
either \texttt{OneLoop}~\cite{oneloop}, 
or \texttt{Golem95C}~\cite{golem95}.
Contributions from top quarks loops are omitted 
as they have been shown to be negligible in the diphoton plus one jet process~\cite{aa1jnlo}.
All ingredients are combined in an automated way
with a numerical phase space integration provided by 
\texttt{MadEvent}~\cite{Maltoni:2002qb}.

\begin{figure}[h]
\begin{center}
\includegraphics[width=6.5cm]{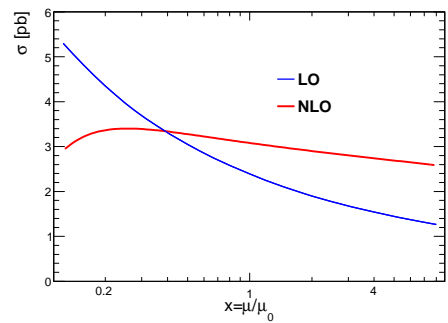} 
\caption{Scale dependence of the total cross section at LO and NLO with $x= \mu/\mu_0$.}
\label{Fig:scalevar}
\end{center}
\end{figure}

The numerical results presented in the following have been calculated at
center-of-mass energy  $\sqrt{s}=8$\,TeV.
For the  jet clustering we used an anti-$k_T$ algorithm~\cite{Cacciari:2008gp} with a cone size
of $R_j=0.5$ provided by 
the {\tt FastJet} package \cite{fastjet}.
We used the CT10 set of parton distributions~\cite{Lai:2010vv} 
as contained in the LHAPDF library~\cite{Whalley:2005nh} and worked with $N_F=5$ massless 
quark flavours.
The following kinematic cuts have been applied:
\begin{eqnarray*}
&&
p_T^{\rm{jet}}>30\mbox{ GeV},\quad  p_T^{\gamma,1}>40 \mbox{~GeV}, \quad
p_T^{\gamma,2}>25\mbox{~GeV},\\
&&|\eta^{\gamma}| \leq 2.5, \quad
|\eta^{j}| \leq 4.7,\quad R_{\gamma ,j} > 0.5,\quad R_{\gamma, \gamma} >0.45.
\end{eqnarray*}
For the photon isolation, we use the smooth cone isolation criterion~\cite{frixione} with 
$R=0.4, n=1$ and $\epsilon=0.05$. Renormalization and factorization scales $\mu$ and $\mu_F$
have been chosen as dynamical scales, 
with the default scale being $\mu_0^2= \frac{1}{4}\,(m_{\gamma\gamma}^2+\sum_j p_{T,j}^2)$, 
and we have used $\mu=\mu_F$.
The behavior of the total cross section when varying 
the scales by a factor of $x\cdot \mu_0$ is shown in
Figure~\ref{Fig:scalevar}. The plot shows a substantial reduction of the scale uncertainty when
including
the NLO corrections. For the total cross section we obtain
$$\sigma_{{\rm LO}}=2.39 ^{+0.66}_{-0.49} \;\rm pb, \; \sigma_{{\rm NLO}}=
3.08^{+0.21}_{-0.18}\; \rm pb\;,$$
where the perturbative uncertainty is estimated by varying $x\in [0.5,2]$. 

\begin{figure}[t]
\begin{center}
\includegraphics[width=7.5cm]{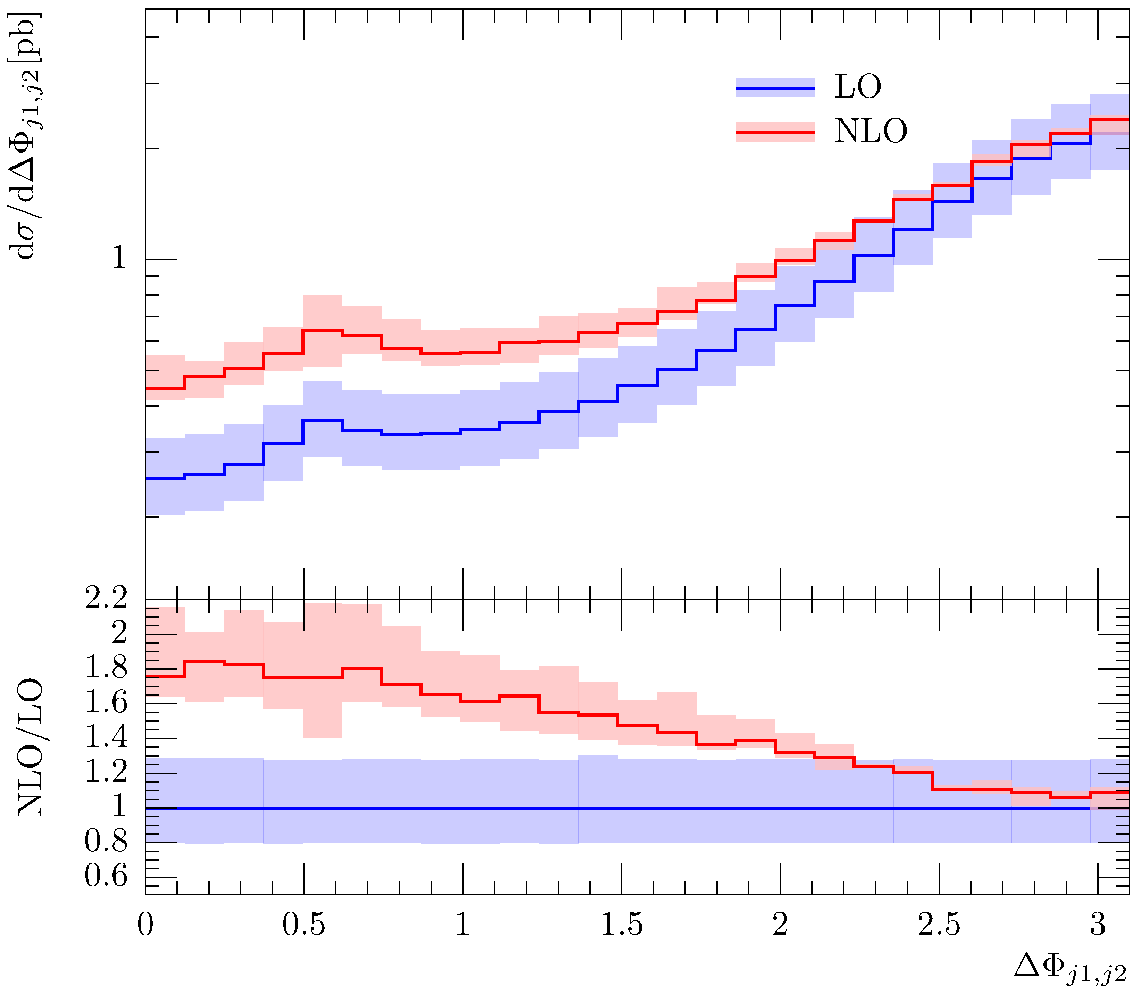} 
\caption{Azimuthal angle $\Delta \Phi(j_1,j_2)$ distribution between the two hardest jets.\label{Fig:dphi}\\}
\includegraphics[width=7.5cm]{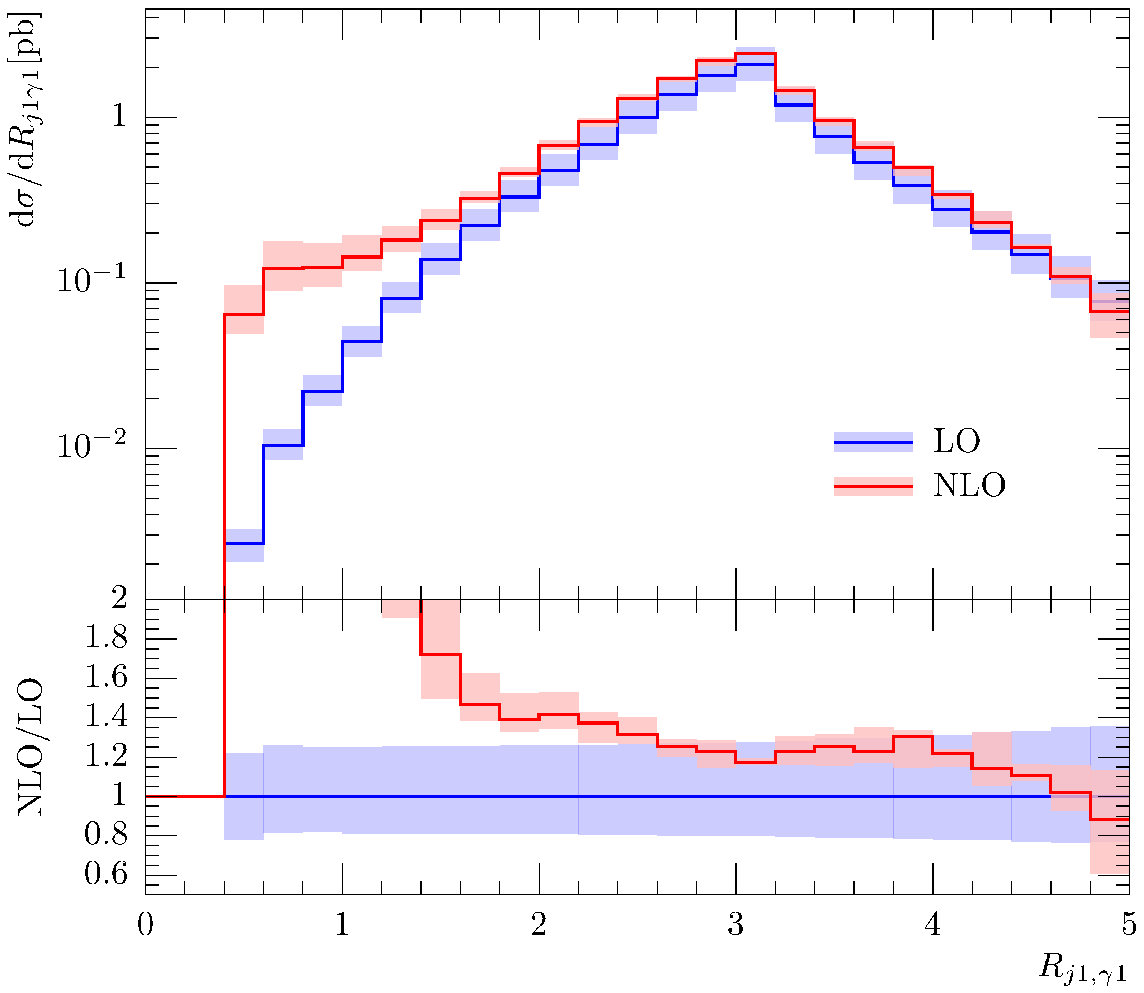} 
\caption{$R$-separation $R(j_1,\gamma_1)$ between the hardest jet and the hardest photon.}
\label{Fig:Rj1a1}
\end{center}
\end{figure}

In Higgs production via vector boson fusion (VBF) and Higgs decay into two photons, 
the CP-properties of the Higgs coupling to electroweak gauge bosons 
are reflected in the azimuthal
distribution of the two tagging jets~\cite{Plehn:2001nj}. 
Fig. \ref{Fig:dphi} shows the distribution of the 
corresponding QCD background for the production of two jets and two photons.
The error band denotes the theoretical uncertainty estimated 
from the variation of renormalization
and factorization scales from $0.5\cdot \mu_0$ to $2\cdot \mu_0$.
Apart from the expected reduction of the scale uncertainty, Figure \ref{Fig:dphi} exhibits a
significant change
of the NLO shape compared to the tree-level calculation. 
This behavior can be understood from the fact that the additional parton
present in the NLO real radiation part enhances the configurations where the jets are 
close in azimuthal angle.
Therefore it is crucial that NLO corrections
are taken into account for a precise estimation of the background.

The $R$-separation between the hardest jet and the hardest photon as shown in Figure \ref{Fig:Rj1a1}
exhibits large differences in the shape at NLO for small values of the $R$-separation. 
This can be explained by the underlying kinematics. At LO,
the pair of hard photon and hard jet being close in $R$-space would have to be counterbalanced with the
soft photon and the soft jet. This is kinematically impossible at LO but 
appears at NLO due to the additional radiation.

\begin{figure}[t]
\begin{center}
\includegraphics[width=8cm]{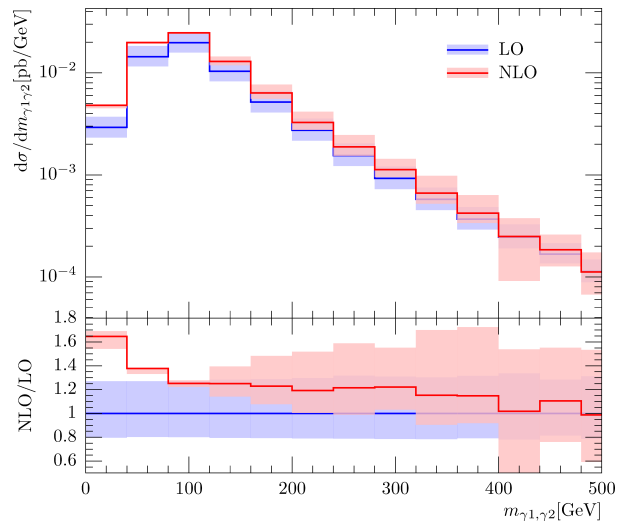} 
\caption{Invariant mass $m_{\gamma \gamma}$ of the two photons.}
\label{Fig:mgg}
\end{center}
\end{figure}

Figure \ref{Fig:mgg} shows the invariant mass of 
the diphoton system. Here as well, inclusion of the NLO corrections 
results in a kinematics-dependent, non-constant correction factor. 
In particular, NLO corrections lead to an enhancement in the low $m_{\gamma \gamma}$ regime. 
This behavior can be explained by the larger final-state 
phase space available at NLO. 
The increase of the 
perturbative uncertainty in the high-mass tail can be understood to be due to the growing 
relative importance of three-jet final states in this region, thereby resulting in a leading order 
scale dependence. 

The substantial corrections to the shapes that we 
observed in several different kinematic distributions are well beyond the 
estimated uncertainties obtained at leading order. They  highlight the importance of genuine NLO 
QCD effects in photon-pair-plus-two-jet production. To obtain background estimates in 
Higgs boson studies from the candidate-plus-two-jet events sample, 
multi-differential distributions are required. Those can often not be extracted reliably 
from sideband data due to lack of statistics, and our calculation provides for the first time 
a sound theoretical prediction for the QCD induced background  
in the highly important two-photon channel. 
With the cuts as defined above, we observe that the total cross section can be predicted at NLO to an 
accuracy of about 10\%. 
At present, our calculation is restricted to 
the smooth cone isolation, which 
eliminates photon-fragmentation contributions and therefore
differs from the fixed-cone isolation prescription 
typically used in experimental studies. 
A more detailed phenomenological study of higher-order QCD effects
in diphoton-plus-two-jet production  
and of the impact of photon isolation issues
will be presented in a subsequent paper.

We thank the other members of the GoSam collaboration for useful discussions.
N.G. and G.H. want to thank Thomas Hahn for technical support of the computing resources,
Gionata Luisoni for his help with Rivet and Sherpa, and the University of Zurich for kind hospitality
while parts of this project were carried out.
This work was supported in part by the Schweizer Nationalfonds under grant
200020-138206, and by the Research Executive Agency (REA) of the
European Union under the Grant Agreement number PITN-GA-2010-264564
(LHCPhenoNet). We acknowledge use of the computing resources of the Rechenzentrum Garching.


\begin{thebibliography}{99}

\bibitem{higgsLHC}
G.~Aad {\it et al.}  [ATLAS Collaboration],
  Phys.\ Lett.\ B {\bf 716} (2012) 1
  [arXiv:1207.7214];
  S.~Chatrchyan {\it et al.}  [CMS Collaboration],
  Phys.\ Lett.\ B {\bf 716} (2012) 30
  [arXiv:1207.7235].

\bibitem{ggHnnlo}
C.~Anastasiou, K.~Melnikov and F.~Petriello,
  Nucl.\ Phys.\ B {\bf 724} (2005) 197
  [hep-ph/0501130];
 M.~Grazzini,
  JHEP {\bf 0802} (2008) 043
  [arXiv:0801.3232].

\bibitem{vhnnlo}
 G.~Ferrera, M.~Grazzini and F.~Tramontano,
  Phys.\ Rev.\ Lett.\  {\bf 107} (2011) 152003
  [arXiv:1107.1164].

\bibitem{vbfnlo}
   T.~Figy, C.~Oleari and D.~Zeppenfeld,
  Phys.\ Rev.\ D {\bf 68} (2003) 073005
  [hep-ph/0306109];
 K.~Arnold, {\it et al.},
  Comput.\ Phys.\ Commun.\  {\bf 180} (2009) 1661
  [arXiv:0811.4559];
  F.~Campanario, T.~M.~Figy, S.~Pl{\"a}tzer and M.~Sj{\"o}dahl,
  arXiv:1308.2932.
\bibitem{ttHnlo}
 W.~Beenakker, S.~Dittmaier, M.~Kramer, B.~Plumper, M.~Spira and P.~M.~Zerwas,
  Nucl.\ Phys.\ B {\bf 653} (2003) 151
  [hep-ph/0211352];
  S.~Dawson, C.~Jackson, L.~H.~Orr, L.~Reina and D.~Wackeroth,
  Phys.\ Rev.\ D {\bf 68} (2003) 034022
  [hep-ph/0305087];
  R.~Frederix, S.~Frixione, V.~Hirschi, F.~Maltoni, R.~Pittau and P.~Torrielli,
  Phys.\ Lett.\ B {\bf 701} (2011) 427
  [arXiv:1104.5613].
\bibitem{ggH1jnlo}
D.~de Florian, M.~Grazzini, Z.~Kunszt,
  Phys.\ Rev.\ Lett.\  {\bf 82} (1999)  5209
  [hep-ph/9902483];
 V.~Ravindran, J.~Smith, W.~L.~Van Neerven,
  Nucl.\ Phys.\  {\bf B634} (2002)  247
  [hep-ph/0201114].

\bibitem{ggH2jnlo}
J.~M.~Campbell, R.~K.~Ellis, G.~Zanderighi,
  JHEP {\bf 0610}, 028 (2006).
  [hep-ph/0608194];\\
 J.~M.~Campbell, R.~K.~Ellis, C.~Williams,
  Phys.\ Rev.\  {\bf D81} (2010)  074023.
  [arXiv:1001.4495].
 H.~van Deurzen, N.~Greiner, G.~Luisoni, P.~Mastrolia, E.~Mirabella, G.~Ossola, T.~Peraro and J.~F.~von Soden-Fraunhofen {\it et al.},
  Phys.\ Lett.\ B {\bf 721}, 74 (2013)
  [arXiv:1301.0493].



\bibitem{ggH3jnlo}
 G.~Cullen, H.~van Deurzen, N.~Greiner, G.~Luisoni, P.~Mastrolia, E.~Mirabella, G.~Ossola,
  T.~Peraro and F.\ Tramontano,
  arXiv:1307.4737.
  
  \bibitem{atlashaa}
 [ATLAS Collaboration],
  ATLAS-CONF-2013-012.

\bibitem{cmshaa}
 [CMS Collaboration],
  CMS-PAS-HIG-13-001.


  
\bibitem{haa}
  J.~R.~Ellis, M.~K.~Gaillard and D.~V.~Nanopoulos,
  Nucl.\ Phys.\ B {\bf 106} (1976) 292;
    M.~A.~Shifman, A.~I.~Vainshtein, M.~B.~Voloshin and V.~I.~Zakharov,
  Sov.\ J.\ Nucl.\ Phys.\  {\bf 30} (1979) 711.
  
\bibitem{ggaannlo}
 S.~Catani, L.~Cieri, D.~de Florian, G.~Ferrera and M.~Grazzini,
  Phys.\ Rev.\ Lett.\  {\bf 108} (2012) 072001
  [arXiv:1110.2375].

\bibitem{nagy}
  V.~Del Duca, F.~Maltoni, Z.~Nagy and Z.~Trocsanyi,
  JHEP {\bf 0304} (2003) 059
  [hep-ph/0303012].


\bibitem{aa1jnlo}
 T.~Gehrmann, N.~Greiner and G.~Heinrich,
  JHEP {\bf 1306} (2013) 058
  [arXiv:1303.0824].

\bibitem{walsh}
K.~Koller, T.~F.~Walsh and P.~M.~Zerwas,
  Z.\ Phys.\ C {\bf 2} (1979) 197.
  
  \bibitem{morgan}
   E.~W.~N.~Glover and A.~G.~Morgan,
  Z.\ Phys.\ C {\bf 62} (1994) 311;
  A.~Gehrmann-De Ridder and E.~W.~N.~Glover,
  Nucl.\ Phys.\ B {\bf 517} (1998) 269
  [hep-ph/9707224].
  
  \bibitem{aleph}
  D.~Buskulic {\it et al.}  [ALEPH Collaboration],
  Z.\ Phys.\ C {\bf 69} (1996) 365;
 A.~Gehrmann-De Ridder, T.~Gehrmann and E.W.N.~Glover,
  Phys.\ Lett.\ B {\bf 414} (1997) 354
  [hep-ph/9705305].
  
  \bibitem{frixione}
   S.~Frixione,
  Phys.\ Lett.\ B {\bf 429} (1998) 369
  [hep-ph/9801442].
  
  
\bibitem{mg4} 
  T.~Stelzer and W.~F.~Long,
  Comput.\ Phys.\ Commun.\  {\bf 81}, 357 (1994)
  [hep-ph/9401258];
J.~Alwall {\it et al.},
  JHEP {\bf 0709}, 028 (2007)
  [arXiv:0706.2334].
  
\bibitem{maddipole} 
  R.~Frederix, T.~Gehrmann and N.~Greiner,
  JHEP {\bf 0809}, 122 (2008)
  [arXiv:0808.2128];
  JHEP {\bf 1006}, 086 (2010)
  [arXiv:1004.2905].
  
\bibitem{Catani:1996vz} 
  S.~Catani and M.~H.~Seymour,
  Nucl.\ Phys.\ B {\bf 485}, 291 (1997)
  [Erratum-ibid.\ B {\bf 510}, 503 (1998)]
  [hep-ph/9605323].
  
  \bibitem{sherpa} 
  T.~Gleisberg, S.~Hoeche, F.~Krauss, M.~Schonherr, S.~Schumann, F.~Siegert and J.~Winter,
  JHEP {\bf 0902}, 007 (2009)
  [arXiv:0811.4622].

\bibitem{gosam} 
  G.~Cullen, N.~Greiner, G.~Heinrich, G.~Luisoni, P.~Mastrolia, G.~Ossola, T.~Reiter and F.~Tramontano,
  Eur.\ Phys.\ J.\ C {\bf 72}, 1889 (2012)
  [arXiv:1111.2034].
  
\bibitem{qgraf} 
  P.~Nogueira,
  J.\ Comput.\ Phys.\  {\bf 105}, 279 (1993).
  
\bibitem{form} 
  J.A.M.~Vermaseren,
  math-ph/0010025.
J.~Kuipers, T.~Ueda, J.A.M.~Vermaseren and J.~Vollinga,
  Comput.\ Phys.\ Commun.\  {\bf 184}, 1453 (2013)
  [arXiv:1203.6543].
\bibitem{spinney} 
  G.~Cullen, M.~Koch-Janusz and T.~Reiter,
  Comput.\ Phys.\ Commun.\  {\bf 182}, 2368 (2011)
  [arXiv:1008.0803].
  
\bibitem{haggies} 
  T.~Reiter,
  Comput.\ Phys.\ Commun.\  {\bf 181}, 1301 (2010)
  [arXiv:0907.3714].
  
  


\bibitem{samurai} 
  P.~Mastrolia, G.~Ossola, T.~Reiter and F.~Tramontano,
  JHEP {\bf 1008}, 080 (2010)
  [arXiv:1006.0710].
  
\bibitem{unitarity} 
  R.~K.~Ellis, W.T.~Giele and Z.~Kunszt,
  JHEP {\bf 0803}, 003 (2008)
  [arXiv:0708.2398].
 G.~Ossola, C.G.~Papadopoulos and R.~Pittau,
  Nucl.\ Phys.\ B {\bf 763}, 147 (2007)
  [hep-ph/0609007].
   P.~Mastrolia, G.~Ossola, C.G.~Papadopoulos and R.~Pittau,
  JHEP {\bf 0806}, 030 (2008)
  [arXiv:0803.3964].
 G.~Ossola, C.G.~Papadopoulos and R.~Pittau,
  JHEP {\bf 0805}, 004 (2008)
  [arXiv:0802.1876].
 G.~Heinrich, G.~Ossola, T.~Reiter and F.~Tramontano,
  JHEP {\bf 1010}, 105 (2010)
  [arXiv:1008.2441].

\bibitem{golem95} 
  T.~Binoth, J.P.~Guillet, G.~Heinrich, E.~Pilon and T.~Reiter,
  Comput.\ Phys.\ Commun.\  {\bf 180}, 2317 (2009)
  [arXiv:0810.0992].
G.~Cullen, J.P.~Guillet, G.~Heinrich, T.~Kleinschmidt, E.~Pilon, T.~Reiter and M.~Rodgers,
  Comput.\ Phys.\ Commun.\  {\bf 182}, 2276 (2011)
  [arXiv:1101.5595].

  
\bibitem{oneloop} 
  A.~van Hameren,
  Comput.\ Phys.\ Commun.\  {\bf 182}, 2427 (2011)
  [arXiv:1007.4716].

  
\bibitem{Maltoni:2002qb} 
  F.~Maltoni and T.~Stelzer,
  JHEP {\bf 0302}, 027 (2003)
  [hep-ph/0208156].
  
\bibitem{Cacciari:2008gp} 
  M.~Cacciari, G.~P.~Salam and G.~Soyez,
  JHEP {\bf 0804}, 063 (2008)
  [arXiv:0802.1189].
  
\bibitem{fastjet} 
  M.~Cacciari and G.~P.~Salam,
  Phys.\ Lett.\ B {\bf 641}, 57 (2006)
  [hep-ph/0512210].
 M.~Cacciari, G.~P.~Salam and G.~Soyez,
  Eur.\ Phys.\ J.\ C {\bf 72}, 1896 (2012)
  [arXiv:1111.6097].
  
\bibitem{Lai:2010vv} 
  H.~-L.~Lai {\it et al.},
  Phys.\ Rev.\ D {\bf 82}, 074024 (2010)
  [arXiv:1007.2241].
  
  
\bibitem{Whalley:2005nh} 
  M.R.~Whalley, D.~Bourilkov and R.C.~Group,
  hep-ph/0508110.
  
\bibitem{Plehn:2001nj} 
  T.~Plehn, D.~L.~Rainwater and D.~Zeppenfeld,
  Phys.\ Rev.\ Lett.\  {\bf 88}, 051801 (2002)
  [hep-ph/0105325].
  
  
  
\end{thebibliography}
\end{document}